\documentstyle[prd,aps,eqsecnum,epsfig]{revtex}
\tighten
\newcommand{\be}{\begin{equation}}
\newcommand{\ee}{\end{equation}}
\newcommand{\Be}{\begin{eqnarray}}
\newcommand{\Ee}{\end{eqnarray}}
\newcommand{\f}{\frac}
\begin{document}
\draft \twocolumn[\hsize\textwidth\columnwidth\hsize\csname
@twocolumnfalse\endcsname

\title{Scattering problem of scalar wave in wormhole geometry}
\author{
Sung-Won Kim\footnote{E-mail address: sungwon@mm.ewha.ac.kr}}
\address{Department of Science Education,
Ewha Women's University, Seoul 120-750, Korea}

\date{\today}
\maketitle

\begin{abstract}
In this paper, we study the scattering problem of the scalar wave
in the traversable Lorentzian wormhole geometry. The potentials
and Schr\"odinger-like equations are found in cases of the static
uncharged and the charged wormholes. The differential scattering
cross sections are determined by the phase shift of the asymptotic
wave function in low frequency limit. It is also found that the
cross section for charged wormhole is smaller than that for
uncharged case by the reduction of the throat size due to the
charge effect.

\end{abstract}

\vskip 1cm \pacs{04.20.Jb, 03.45.Pm, 03.65.Nk}

\vskip2pc]

\bigskip
\section{Introduction}

The wormhole has the structure which is given by two
asymptotically flat regions and a bridge connecting two
regions\cite{MT88}. For the Lorentzian wormhole to be traversable,
it requires exotic matter which violates the known energy
conditions. To find the reasonable models, there had been studying
on the generalized models of the wormhole with other matters
and/or in various geometries. Among the models, the matter or wave
in the wormhole geometry and its effect such as radiation are very
interesting to us. The scalar field could be considered in the
wormhole geometry as the primary and auxiliary effects\cite{K00}.
Recently the solution for the electrically charged case was also
found \cite{KL01}. Scalar wave solutions in the wormhole
geometry\cite{KSB94,KMMS95} was in special wormhole model only the
transmission and reflection coefficients were found.  The
electromagnetic wave in wormhole geometry is recently
discussed\cite{BH00} along the method of scalar field case. These
wave equations in wormhole geometry draws attention to the
research on radiation and wave. Also there was a suggestion that
the wormhole would be one of the candidates of the gamma ray
bursts\cite{TRA98}.

For the gravitational radiation in any forms, the scattering
problem to calculate the cross section in more generalized models
of wormhole should be considered. Thus the study of scalar,
electromagnetic, gravitational waves in wormhole geometry is
necessary to the research on the gravitational radiation.

In this paper, we study the scalar wave in static uncharged and
charged wormhole and find the differential scattering cross
sections. We also compare the results each other in both cases to
see the charge effect on scattering problem. Here we adopt the
geometrical unit, {\it i.e.}, $G=c=\hbar=1$.

\vskip .5cm

${}^*$E-mail address: sungwon@mm.ewha.ac.kr

\section{Static Uncharged Wormhole}

The spacetime metric for static uncharged wormhole is given as \be
ds^2 = -e^{2\Lambda(r)}dt^2 + \f{dr^2}{1-b(r)/r} + r^2
(d\theta^2+\sin^2\theta d\phi^2), \ee where $\Lambda(r)$ is the
lapse function and $b(r)$ is the wormhole shape function. They
are assumed to be dependent on $r$ only for static case.

The wave equation of the minimally coupled massless scalar field
is given by
\be
\Box\Phi =
\f{1}{\sqrt{-g}}
\partial_\mu
(\sqrt{-g} g^{\mu\nu}\partial_\nu \Phi ) = 0. \ee In spherically
symmetric space-time, the scalar field can be separated by
variables, \be \Phi_{lm} = Y_{lm}(\theta, \phi)\f{u_l(r,t)}{r},
\label{eq:def1} \ee where $Y_{lm}(\theta, \phi)$ is the spherical
harmonics and $l$ is the quantum angular momentum.

If $l=0$ and the scalar field $\Phi(r)$ depends on $r$ only, the
wave equation simply becomes the following relation\cite{KK98}:
\be e^\Lambda \sqrt{1-\f{b}{r}} r^2 \f{\partial}{\partial r}\Phi =
A = \mbox{const.} \ee In this relation, the back reaction of the
scalar wave on the wormhole geometry is neglected. Thus the static
scalar wave without propagation is easily found as the integral
form of \be \Phi = A \int e^{-\Lambda} r^{-2} \left( 1- \f{b}{r}
\right)^{-1/2} dr. \ee The scalar wave solution was already given
to us for the special case of wormhole in Ref.~\cite{KL01,KK98}.

More generally, if the scalar field $\Phi$ depends
on $r$ and $t$, the wave equation after the
separation of variables $(\theta,\phi)$ becomes
\be
- \ddot{u}_l + \f{\partial^2 u_l}{\partial r_*^2 } =
V_l \,\,u_l,
\ee
where the potential is
\Be
V_l(r)&=&\f{L^2}{r^2}e^{2\Lambda} + \f{1}{r}e^\Lambda
\sqrt{1-\f{b}{r}}\f{\partial}{\partial r}\left(e^\Lambda
\sqrt{1-\f{b}{r}}\right) \nonumber \\
&=& e^{2\Lambda}\left( \f{l(l+1)}{r^2} - \f{b'r-b}{2r^3} +
\f{1}{r}\left(1-\f{b}{r}\right)\Lambda' \right) \Ee and the proper
distance $r_*$ has the following relation to $r$: \be
\f{\partial}{\partial r_*} = e^\Lambda \sqrt{1-\f{b}{r}}
\f{\partial}{\partial r}. \ee Here, $L^2=l(l+1)$ is the square of
the angular momentum.

The properties of the potential are determined by the shape of it,
if only the explicit forms of $\Lambda$ and $b$ are given. If the
time dependence of the wave is harmonic as $u_l(r,t) =
\hat{u}_l(r,\omega)e^{-i\omega t} $, the equation becomes \be
\left( \f{d^2}{dr^2_*} + \omega^2 - V_l(r)
\right)\hat{u}_l(r,\omega) = 0. \ee It is just the Schr\"odinger
equation with energy $\omega^2$ and potential $V_l(r)$. When
$e^{2\Lambda}$ is finite, $ V_l $ approaches zero as $r
\rightarrow \infty $, which means that the solution has the form
of the plane wave $ \hat{u}_l \sim e^{\pm i \omega r_*}$
asymptotically. The result shows that if a scalar wave passes
through the wormhole the solution is changed from $e^{\pm i \omega
r}$ into $e^{\pm i \omega r_*}$, which means that the potential
affects the wave and experience the scattering. As $ r \rightarrow
b$(near throat), the potential has the finite value of $ V_l
\simeq e^{2\Lambda(b)} \f{l(l+1)+1/2}{b^2}$.

As the simplest example for this problem, we consider the special
case $ (\Lambda = 0, b = b_0^2/r)$ as usual, the potential should
be in terms of $r$ or $r_*$ as \be V_l = \f{l(l+1)}{r^2} +
\f{b_0^2}{r^4} ~\stackrel{\mbox{or}}{=}~ \f{l(l+1)}{r_*^2+b_0^2} +
\f{b_0^2}{(r_*^2+b_0^2)^2}, \label{eq:pot} \ee where the proper
distance $r_*$ is given by \be r_* = \int
\f{1}{\sqrt{1-b_0^2/r^2}}dr = \sqrt{r^2-b_0^2}. \label{eq:prop1}
\ee This is the hyperbolic relation between $r_*$ and $r$ which is
plotted in Fig.~4. The potentials are depicted in Fig.~1. The
potential has the maximum value as \be V_l(r)|_{\rm max} =
V_l(r_*)|_{\rm max} = V_l(b_0) = \f{l(l+1)+1}{b_0^2}. \ee

\begin{figure}
\begin{center}
\psfig{figure=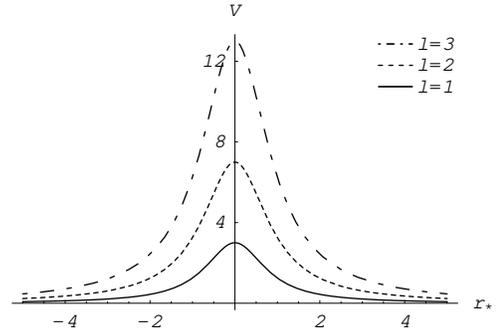,width=3.0in,angle=0} \caption{Plot of the
potentials of the specified wormhole $b=b_0^2/r$ for $l=1,2,3$.
Here we set $b_0=1$. }
\end{center}
\end{figure}

Now we can see the characteristics of the potential
Eq.~(\ref{eq:pot}) in the specified limits. When $l=0$, the
potential and its asymptotic value are \be V_l(r_*) =
\f{b_0^2}{(r_*^2+b_0^2)^2}~ \stackrel{
r_*\rightarrow\infty}{\longrightarrow}~\f{b_0^2}{r_*^4}. \ee When
$l\gg 1 $, the potential and its asymptotic value are \be V_l(r_*)
\approx \f{l^2}{r_*^2+b_0^2}~ \stackrel{
r_*\rightarrow\infty}{\longrightarrow}~\f{l^2}{r_*^2}. \ee As $r_*
\rightarrow \infty (r_*^2 \gg b_0^2)$ independently of the value
of $l$, the potential has the asymptotic value as \be V_l \simeq
\f{l(l+1)}{r_*^2} - \f{[l(l+1)-1]b_0^2}{r_*^4}. \ee

From the potential Eq.~(\ref{eq:pot}), the transmission
coefficient can be calculated by WKB approximation as \be
|T|_{(l\omega)}^2 = \exp \left( - 2\int_{a_-}^{a_+} \left(
\f{l(l+1)}{r_*^2+b_0^2} + \f{b_0^2}{(r_*^2+b_0^2)^2} - \omega^2
\right)^{1/2} dr_* \right), \ee where the upper and lower
integration limits $a_+$ and $a_-$ are \be a_\pm = \pm \left(
-b_0^2 +
\f{l(l+1)}{2\omega^2}+\f{\sqrt{l^2(l+1)^2+4b_0^2\omega^2}}{2\omega^2}
\right)^{1/2}. \ee The transmission coefficient means the
probability that the scalar wave can pass through the throat of
the wormhole, even though it does not have enough energy to
overcome the potential. A few transmission coefficients are
depicted in Fig.~2.

For large $l$, it becomes \be |T|^2 \approx \left(
\f{\sqrt{l^2+\omega^2b_0^2}-l}{\sqrt{l^2+\omega^2b_0^2}+l}
\right)^{2l}. \ee

\begin{figure}
\begin{center}
\psfig{figure=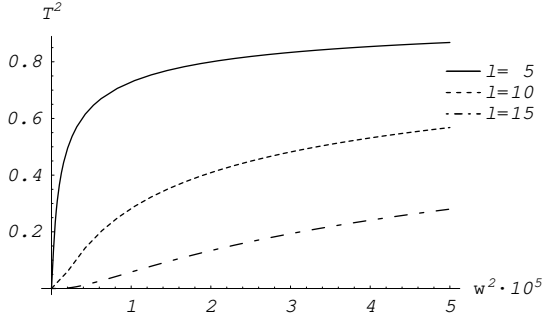,width=3.0in,angle=0} \caption{The
$\omega$-dependency of $|T|_{(l\omega)}^2$ for the specified
wormhole $b=b_0^2/r$ for $l=5,10,15$. Here we set $b_0=1$. }
\end{center}
\end{figure}

\noindent When $l=0$, it is exactly given as \be |T|^2 =
e^{-A(\omega)}, \ee where \Be && A(\omega) \nonumber\\ &&=
\f{b_0^2\pi}{\omega^2}\left( \f{2}{b_0} -
\f{(\f{\omega}{b_0})^{1/2}\sqrt{2\pi}
{}_3F_2[\{-\f{1}{4},\f{1}{4},\f{3}{4}\},\{\f{1}{2},\f{5}{4}\};
b_0^2\omega^2]}{\Gamma(\f{1}{4})\Gamma(\f{5}{4})} \right. \nonumber \\
&& \quad \left. + \f{(\f{\omega}{b_0})^{3/2}b_0^2\sqrt{2\pi}
{}_3F_2[\{\f{1}{4},\f{3}{4},\f{5}{4}\},\{\f{3}{2},\f{7}{4}\};
b_0^2\omega^2]}{\Gamma(-\f{1}{4})\Gamma(\f{7}{4})}\right).
 \Ee
Here ${}_pF_q(\{a_1,\cdots,a_p\},\{b_1,\cdots,b_q\};z)$ is the
generalized hypergeometric function.

For more deep understanding on scattering problem, we should find
the differential cross section. Since we discuss the low frequency
case similar to the black hole case\cite{FHM88}, we approach the
problem with the different definition of $\Phi$ by using $R(r,t)$
instead of $u_l(r,t)/r$ unlike Eq.~(\ref{eq:def1}) as \be \Phi =
R(r,t)Y(\theta,\phi). \ee The equation, in terms of $R$, becomes
\Be && -e^{-2\Lambda}r^2 \ddot{R} + e^{-\Lambda}
\sqrt{1-\f{b}{r}}\f{\partial}{\partial r} \left(
e^\Lambda\sqrt{1-\f{b}{r}}~r^2 \f{\partial}{\partial r}R \right)
\nonumber \\ && \qquad = l(l+1)R. \Ee The equation can be
rewritten as \be \f{d}{dr}\Delta\f{dR}{dr}+\left(
\f{(r^2\omega^2)^2}{\Delta}-l(l+1) \f{e^\Lambda}{\sqrt{1-b/r}}
\right)R = 0, \label{eq:wave} \ee where $ \Delta=e^\Lambda
\sqrt{1-\f{b}{r}}r^2 $. If we set $ R = \Delta^{-1/2}u $, the
first term of the equation Eq.~(\ref{eq:wave}) should be \Be
\f{d}{dr}\Delta\f{dR}{dr} &=& \Delta^{1/2} \left\{ \f{d^2u}{dr^2}
+ \left( \f{1}{4}\f{1}{\Delta^2} \left( \f{d\Delta}{dr} \right)^2
\right. \right.\nonumber \\  &&\qquad  \left. \left.
-\f{1}{2}\f{1}{\Delta}\f{d^2\Delta}{dr^2} \right) u \right\}. \Ee
Thus the equation Eq.~(\ref{eq:wave}) becomes as \Be
\f{d^2u}{dr^2} &+& \left( \f{(r^2\omega^2)^2}{\Delta}-l(l+1)
\f{e^\Lambda}{\sqrt{1-b/r}} + \f{1}{4}\f{1}{\Delta} \left(
\f{d\Delta}{dr} \right)^2 \right. \nonumber \\
&& \qquad  \left. - \f{1}{2}\f{d^2\Delta}{dr^2} \right)
\f{u}{\Delta} = 0. \label{eq:wave1}\Ee

For example, $ \Lambda=0, ~ b= b_0^2/r$ like the potential
problem, then $\Delta$ should be \be \Delta = \sqrt{r^4-b_0^2r^2}.
\ee The equation Eq.~(\ref{eq:wave1}) is expanded in power of $r$
as \Be \f{d^2u}{dr^2} &+& \left( \omega^2 +
\f{\omega^2b_0^2-l(l+1)}{r^2} + b_0^2~ \f{\omega^2b_0^2+\f{1}{2} -
l(l+1)}{r^4}\right. \nonumber\\
&&\qquad\qquad \left.+ O(r^{-6}) \right) u ~\simeq~ 0.
\label{eq:exp} \Ee As $r \rightarrow \infty$ neglecting the terms
$O(r^{-4})$, the equation becomes \be \f{d^2u}{dr^2} + \left(
\omega^2 + \f{\omega^2b_0^2}{r^2}-\f{l(l+1)}{r^2} \right)u
~\simeq~ 0. \label{eq:wave2} \ee When the frequency is enough low
for being as $ \omega^2 b_0^2 < l(l+1)$, the scattering is elastic
and the solution to the equation is \be u_{l\omega} = r
j_\lambda(\omega r). \label{eq:sol} \ee Here $\lambda(\lambda+1) =
l(l+1)-\omega^2b_0^2$ or $\lambda =
-\f{1}{2}+\sqrt{(l+\f{1}{2})^2-\omega^2b_0^2}$ and $j_\lambda(r)$
is the spherical Bessel function. The turning point is \be r_{\rm
TP} = \sqrt{\f{l(l+1)}{\omega^2}-b_0^2}. \ee When $\omega^2b_0^2
\ge l(l+1)$, there is a total absorption and no turning point.
Since the solution Eq.~(\ref{eq:sol}) asymptotically becomes \be
u_{l\omega} \sim \f{1}{\omega}\sin (\omega r - \f{\pi}{2}\lambda),
\ee  the phase shift of the wave function by the potential is \be
\delta_l = -\f{\pi}{2}\left( \sqrt{(l+\f{1}{2})^2-\omega^2b_0^2} -
\left( l + \f{1}{2}  \right) \right). \label{eq:delta1} \ee The
phase shift $\delta_l$ is obviously positive, which means that the
potential is attractive. The scattering amplitude in partial wave
expansion is \be f(\theta) = \f{1}{\omega}\sum_{l=0}^\infty (2l+1)
P_l(\cos\theta) e^{i\delta_l}\sin\delta_l. \ee If $\omega^2b_0^2$
and $\delta_l$ are very small, the scattering amplitude is given
by \Be f(\theta) &\simeq& \f{1}{\omega}\sum_{l=0}^\infty (2l+1)
P_l(\cos\theta) \delta_l
\nonumber \\
&\simeq& \f{\pi\omega b_0^2}{2} \sum_{l=0}^\infty P_l(\cos\theta)
\nonumber \\
&=& \f{\pi\omega b_0^2}{4}\f{1}{\sin\f{\theta}{2}}. \Ee The
relation \be \delta_l \simeq \f{\pi\omega^2 b_0^2}{2(2l+1)}
\label{eq:delta2} \ee is used for the second line. The equation
(\ref{eq:delta2}) is derived from Eq.~(\ref{eq:delta1}) by using
the fact that $\omega^2b_o^2 \ll 1$. Therefore, the differential
cross section is \be \f{d\sigma}{d\Omega} = |f(\theta)|^2 =
\f{\pi^2\omega^2 b_0^4}{16\sin^2\f{\theta}{2}}.
\label{eq:cross1}\ee

In the case of very low frequency, $\f{l}{b_0\omega}
\rightarrow\infty $ neglecting $b_0\omega$ term, the equation
Eq.~(\ref{eq:wave2}) becomes \be \f{d^2u}{dr^2} + \left( \omega^2
- \f{l(l+1)}{r^2} \right)u ~\simeq~ 0, \ee which means no
scattering, since any term is not appeared except energy
$\omega^2$ and centrifugal term. However, it is Coulomb scattering
for black hole, since there is the term $r^{-1}$\cite{FHM88}. The
solution to the equation for black hole is \be u = j_l(\omega r).
\ee

If we remain the term $r^{-4}$ with low frequency approximation in
Eq.~(\ref{eq:exp}), then \Be \f{d^2u}{dr^2} &+& \left( \omega^2 -
\f{l(l+1)}{r^2} + b_0^2~ \f{1/2 - l(l+1)}{r^4} \right)u
\nonumber\\ && \qquad\qquad~\simeq~ 0 \Ee The equation can be
solved as the problem with an $r^{-4}$ potential. In this case,
the cross section is given in most text books on quantum
mechanics\cite{BOOK}. In WKB approximation, the wave function has
the asymptotic form of \be u_l \sim D\sin\left( \int_{r_0}^r
\sqrt{\omega^2-U-\f{(l+1/2)^2}{r^2}}dr + \f{\pi}{4}\right), \ee
where \be U(r) = \f{b_0^2[l(l+1)-1/2]}{r^4}. \label{eq:pot3}\ee
The phase shift in this wave function should be \Be \delta_l &=&
\lim_{R\rightarrow\infty}\left( \int_a^R
\sqrt{\omega^2-U-\f{(l+1/2)^2}{r^2}}dr \right.\nonumber \\
&& \qquad \qquad \qquad\left. - \int_{a_0}^R
\sqrt{\omega^2-\f{(l+1/2)^2}{r^2}}dr \right) \nonumber \\
&\simeq& - \f{1}{2} \int_{l/\omega}^\infty U(r)\left( \omega^2 -
\f{l^2}{r^2} \right)^{-1/2} = -\f{b_0^2\omega\pi}{4l}, \Ee where
$U$ is assumed to be very small in asymptotic
region comparing
to the two terms in the
square root, even though $l$ is defined as the large value.
Here, the phase shift is negative and it is natural
by the repulsive potential Eq.~(\ref{eq:pot3}).
With this phase shift, the differential cross section
should be \be \f{d\sigma}{d\Omega} \simeq
\f{b_0^4\pi^2}{16\sin^2\f{\theta}{2}}. \ee This is restrictedly
calculated in the limit of
large angular momentum and small $U$. As we see, it is
independent of
$\omega$ and the $\theta$-dependence is the same as
that of the $r^{-2}$ potential case, Eq.~(\ref{eq:cross1}).

\section{Charged Wormhole}

The electrically charged wormhole is given by\cite{KL01} \Be ds^2
&=& -\left( 1 + \f{Q^2}{r^2} \right)dt^2 + \left(1- \f{b(r)}{r} +
\f{Q^2}{r^2} \right)^{-1}dr^2 \nonumber \\
&& \qquad + r^2 (d\theta^2+\sin^2\theta d\phi^2). \Ee When $Q=0$,
this becomes the Morris-Thorne type wormhole spacetime\cite{MT88}
and when $b=0$, this becomes the Reissner-Nordstr\"om black hole
with zero mass. If we replace $b - \f{Q^2}{r}$ by $b_{\rm eff}$ as
\be b~\rightarrow~b_{\rm eff} = b - \f{Q^2}{r}, \ee it
self-consistently satisfies the field equations for the wormhole
without $Q$. When
$b=b_0^{\f{2\beta}{2\beta+1}}r^{\f{1}{2\beta+1}}$, the radius
which limits the range of the wormhole becomes \be r_0 =
Q^{\f{2\beta+1}{\beta+1}}b_0^{\f{\beta}{2\beta+1}}. \ee For the
special case of $\beta = -1$ or $b = b_0^2 /r$, $Q^2 < b_0^2$ is
the condition that is required for maintaining the wormhole under
the addition of the charge.

The metric for the charged case with scalar field is also given
by\cite{KL01} \Be ds^2 &=& -dt^2 + \left(1- \f{b(r)}{r} +
\f{\alpha}{r^2} \right)^{-1}dr^2 \nonumber \\
&& \qquad \qquad+ r^2 (d\theta^2+\sin^2\theta d\phi^2).
\label{eq:scal} \Ee The substitution $b \rightarrow b_{\rm eff} =
b - \f{\alpha}{r}$ does not change the equation like the
electrically charged case and $\alpha < b_0^2$ is also the
condition for being safe under the addition of scalar charge.

To see the scalar wave in the charged wormhole, we first analyze
the properties of the potential. In this case, only the following
substitutions will be enough for the study: \be
e^{2\Lambda}=\left( 1 + \f{Q^2}{r^2} \right) \qquad \mbox{and}
\qquad b~\rightarrow~b_{\rm eff} = b - \f{Q^2}{r} \label{eq:sub}
\ee

For the special case of $b = \f{b_0^2}{r}$, the proper length
will be very complicated form as
\Be
r_* &=& \int
\f{1}{e^\Lambda\sqrt{1-b_{\rm eff}/r}} = \int
\f{r^2dr}{\sqrt{(r^2+Q^2)(r^2+Q^2-b_0^2)}} \nonumber\\
&=&
\sqrt{Q^2-b_0^2}E\left(\sin^{-1}\f{\sqrt{b_0^2-r^2-Q^2}}{b_0}\left|
-\f{b_0^2}{b_0^2-Q^2}\right.\right), \label{eq:prop2} \Ee where
$E(\alpha|\beta)$ is the Elliptic integral of the second kind
defined by \be E(\alpha \mid \beta)=\int_0^\alpha
\sqrt{1-\beta\sin^2\theta}d\theta. \ee With the substitution
Eq.~(\ref{eq:sub}), the potential will be \Be V_l(r)
&=& \left( 1 + \f{Q^2}{r^2} \right)  \f{l(l+1)}{r^2} + \left( 1 +
\f{2Q^2}{r^2} \right) \f{1}{r^4}(b_0^2-Q^2) \nonumber\\
&& \qquad \qquad - \f{Q^2}{r^4}~>~0. \Ee If we examine the proper
length $r_*$ and potential $V_l$ for the limit of infinite
distance ($r\rightarrow\infty$) and near the throat of the
wormhole($r^2\rightarrow b_0^2 - Q^2$), we can figure out the
rough form of the potential. Since \be
\lim_{r\rightarrow\infty}r_* = \pm \infty \qquad \mbox{and} \qquad
\lim_{r^2\rightarrow b_0^2 - Q^2}r_* = 0, \ee the proper length
has the similar form to the uncharged case. At the asymptotic
region, the potential approach \be \lim_{r\rightarrow \infty}V(r),
V(r_*) = 0. \ee Since \be \f{dV}{dr_*} = \f{dV}{dr}\f{dr}{dr_*} =
\f{dV}{dr}\sqrt{\left( 1 + \f{Q^2}{r^2} \right)\left( 1 -
\f{b_0^2}{r^2} + \f{Q^2}{r^2} \right)} \ee  and $\f{dV}{dr_*}=0$
when $\f{dV}{dr} = 0$, the potential at the throat has a finite
common maximum value as \be \lim_{r\rightarrow b_0^2 - Q^2} V(r),
V(r_*) = V(r)|_{\rm max}, V(r_*)|_{\rm max}. \ee The potential
form is compared with the case of $Q^2=0$, so that \Be V_{\rm
max}|_{Q^2\neq 0} &=& \f{l(l+1)}{(b_0^2-Q^2)}\left( 1 +
\f{Q^2}{(b_0^2-Q^2)} \right) + \f{b_0^2}{(b_0^2-Q^2)^2}
\nonumber\\
&& \quad > \f{l(l+1)}{b_0^2} + \f{1}{b_0^2} = V_{\rm
max}|_{Q^2=0}. \label{eq:max} \Ee As we see in Eq.~(\ref{eq:max}),
the charge effect contract the throat size of wormhole and it
causes the maximum of the potential to be increased. Since the
term in the bracket of the following equation is positive for
$l\geq 1$ in general, the potentials of both cases are given as
\Be V|_{Q^2\neq 0} &=& \f{l(l+1)}{r^2} + \f{b_0^2}{r^4} +
\f{Q^2}{r^4}\left( l(l+1) - 2 +
\f{2}{r^2}(b_0^2-Q^2) \right)\nonumber\\
&& \qquad\qquad
> V|_{Q^2=0}. \Ee With these
results, the potentials are plotted in Fig.~3 so that the
potential for $Q^2 \neq 0$ is above the potential for $Q^2 = 0$.
For the relation between $r_*$ and $r$,

\begin{figure}
\begin{center}
\psfig{figure=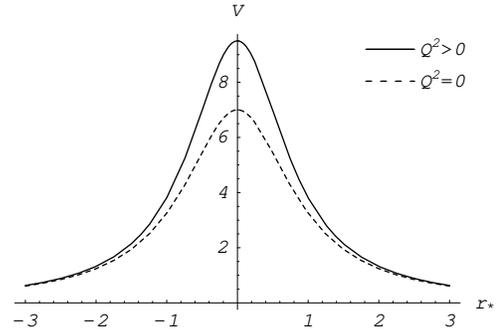,width=3.0in,angle=0} \caption{Potentials
for charged ($Q^2\neq 0$) and uncharged ($Q^2=0$) wormholes. Here
we set $b_0=1$, $Q^2=0.5$, and $l=2$. }
\end{center}
\end{figure}

\begin{figure}
\begin{center}
\psfig{figure=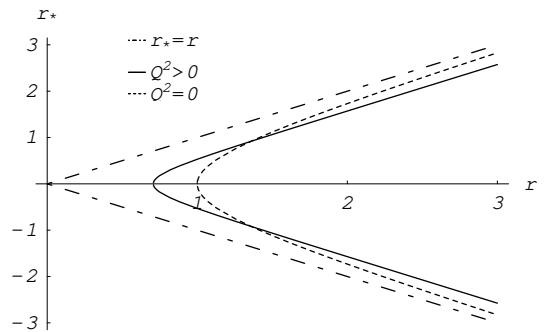,width=3.0in,angle=0} \caption{The relation
between $r$ and $r_*$ for charged($Q^2\neq 0$) and
uncharged($Q^2=0$) wormholes. Here we set $b_0=1$ and $Q^2=0.5$. }
\end{center}
\end{figure}

\Be \f{dr_*}{dr}|_{Q^2\neq 0} &=& \left(1+\f{Q^2}{r^2}\right)^{-1}
\left( 1 - \f{b_0^2}{r^2}+ \f{Q^2}{r^2} \right)^{-1/2} \nonumber\\
&& \quad ~<~ \left( 1 - \f{b_0^2}{r^2}\right)^{-1/2} =
\f{dr_*}{dr}|_{Q^2=0} \Ee  is plotted in Fig. 4. It is based on
the relations of $r_*$ and $r$, {\it i.e.}, Eqs.~(\ref{eq:prop1})
and (\ref{eq:prop2}) for charged and uncharged wormhole. At
asymptotic region, $\lim_{r\rightarrow\infty}r_* = \pm r$ in both
cases.

Since $V_{Q^2\neq 0} > V_{Q^2=0}$ in the form of the potentials,
the transmission coefficient is $|T|^2_{Q^2\neq 0}
< |T|^2_{Q^2=0}$, which
seems that the scalar wave can transmit harder in charged
wormhole than neutral uncharged wormhole.

As $r \rightarrow \infty$ neglecting the terms $O(r^{-4})$, the
equation becomes as \be \f{d^2u}{dr^2} + \left( \omega^2 +
\f{\omega^2(b_0^2-3Q^2)}{r^2}-\f{l(l+1)}{r^2} \right)u ~\simeq~ 0.
\ee Since the $b_0^2$ term is simply replaced by $b_0^2-3Q^2$, the
scattering cross section should be \be \f{d\sigma}{d\Omega} =
|f(\theta)|^2 = \f{\pi^2\omega^2
(b_0^2-3Q^2)^2}{16\sin^2\f{\theta}{2}} <
\f{d\sigma}{d\Omega}|_{Q^2=0}. \ee In case of wormhole with scalar
field, because only $\Lambda=0$ is considered as we see in
Eq.~(\ref{eq:scal}), the term $b_0^2-3Q^2$ for electrically
charged case is just $b_0^2-\alpha$ for scalar field case, so that
the cross section should be \be \f{d\sigma}{d\Omega} =
\f{\pi^2\omega^2 (b_0^2-\alpha)^2}{16\sin^2\f{\theta}{2}} <
\f{d\sigma}{d\Omega}|_{\alpha=0}. \ee Of course, the interaction
of the wave with the scalar field is neglected here. In both
cases, cross sections are reduced by the charge effect.

\section{Discussion}

Here we studied the scattering problem for the scalar wave in
static uncharged and charged wormholes. The interactions of the
scalar wave with the charge are neglected. If we consider them,
the effect would make sense. The scattering cross sections are
found and the charge effects are examined in this scattering
problem. As further research, we will  try other waves for various
wormholes, such as the rotating wormhole and cosmological model
with wormhole.

\acknowledgements

This work was supported by  Korea Research Foundation
Grant(KRF-2000-041-D00128).

\end{document}